 \documentclass[final,5p,times,twocolumn]{elsarticle}
\usepackage{amsmath}

\usepackage{amssymb}

\makeatletter
\def\ps@pprintTitle{%
  \let\@oddhead\@empty
  \let\@evenhead\@empty
  \let\@oddfoot\@empty
  \let\@evenfoot\@oddfoot
}
\makeatother

\journal{Physics Letters B}

\begin{document}

\begin{frontmatter}



\title{Conductivity Tensor in a Holographic Quantum Hall Ferromagnet}

\author{Joel Hutchinson$^a$, Charlotte Kristjansen$^b$  , Gordon W. Semenoff$^a$ }
\address[label1]{Department of Physics and Astronomy, University of British Columbia,\\ 6224 Agricultural Road, Vancouver, British Columbia, Canada V6T 1Z1}
\address[label2]{Niels Bohr Institute, Copenhagen University,\\ Blegdamsvej 17, 2100 Copenhagen \O, Denmark}



\begin{abstract}
The Hall and longitudinal conductivities of a recently studied holographic model of a quantum Hall 
ferromagnet are computed using the Karch-O'Bannon technique. In addition, the
low temperature entropy of the model is determined.
  The holographic model has a phase
transition as the Landau level filling fraction is increased from zero to one.   We argue that this phase
transition allows the longitudinal conductivity to have features 
qualitatively similar to those of two dimensional electron gases in the integer quantum Hall regime.  
The argument also applies to the low temperature limit of the entropy.
The Hall
conductivity is found to have an interesting structure. Even though it does not exhibit Hall plateaux, it has
a flattened dependence on the filling fraction with a jump, analogous to the interpolation
between Hall plateaux, 
at the phase transition.   
\end{abstract}


\end{frontmatter}



Quantum Hall ferromagnetism is an interesting
example of dynamical symmetry breaking.  It  was predicted and observed in two dimensional electron gases formed by 
semiconductor  
heterojunctions  \cite{quantumhallferromagnet0}-\cite{quantumhallferromagnet2} and it has more recently been observed 
in graphene in the integer quantum Hall regime \cite{newhallplateaux}-\cite{newhallplateaux3}.  
When many-body interactions are weak,
this ferromagnetism has a simple mechanism \cite{quantuhallferromagnet3.5}-\cite{quantumhallferromagnet2}\cite{Semenoff:2011ya}.  
For example,  an electron with two  spin states and negligible Zeeman interaction has two-fold degenerate
 Landau levels.  When a 2-fold degenerate level is precisely half-filled,  that is, at filling fraction $\nu=1$,  
the electrons can minimize their Coulomb exchange energy by occupying those states which have only one of the two spin labels.  
The result is  spontaneous 
breaking of spin symmetry and splitting of the degeneracy of the Landau level by the formation of a charge-gapped
incompressible integer quantum Hall state at $\nu=1$.    
A similar mechanism is thought to work for any integer filling fraction in a Landau level with higher
degeneracy.  
Graphene has an emergent SU(4) symmetry which would result in four-fold degenerate Landau levels.
This degeneracy is seen to be completely 
resolved in sufficiently strong magnetic fields. Evidence that the mechanism is  dynamical symmetry breaking  is seen in 
the magnitude of the energy gaps, which are too  large to be accounted for by residual non-symmetric interactions and which 
are  characteristic of the scale of the Coulomb interaction, which is very strong in graphene\cite{kim}.  This raises the 
question as to whether quantum Hall ferromagnetism can be understood at strong coupling.  \footnote{Some work in this direction 
considers
systematic re-summations 
of perturbation theory which have been studied in the closely related framework known as magnetic catalysis of chiral symmetry breaking 
\cite{cat0}-\cite{Shovkovy:2012zn}. It suggests that magnetic catalysis, which is indistinguishable from quantum Hall ferromagnetism
in this particular system, 
can still occur when many-body interactions are appreciable. Spontaneous symmetry breaking in a magnetic field in the charge neutral case
is already well known for the holographic D3-D5 brane system~\cite{Filev:2009xp,Evans:2010hi}.}
 
Recently, a holographic model where quantum Hall ferromagnetism persists in the strong coupling limit has been developed
\cite{Kristjansen:2012ny}-\cite{Kristjansen:2013hma}. 
The model is a D3-probe-D5 brane system which is dual to a super-conformal
 defect field theory with $N_5$ complex fundamental representation 
 hypermultiplets (where $N_5$ is the number of D5 branes) 
 occupying a 2+1 dimensional subspace of 3+1 dimensional
space-time.  The system is Lorentz invariant, which can be regarded as analogous to the emergent Lorentz symmetry of graphene \cite{Semenoff:1984dq}. 
The 3+1-dimensional bulk contains ${\mathcal N}=4$ supersymmetric Yang-Mills theory with gauge group SU(N).
This theory is readily studied in the large $N$ planar limit and the probe limit where $N_5\ll N$. 
The conformal field theory has a tuneable dimensionless coupling constant, the 't Hooft coupling $\lambda=g_{\rm YM}^2N$
of the ${\mathcal N}=4$ Yang-Mills theory.  

One can introduce a non-zero temperature and a U(1) charge density and constant external magnetic field 
for the hypermultiplets.
These deformations break supersymmetry.  Moreover, 
in the limit of weak coupling, $\lambda\ll1$, as discussed in \cite{Kristjansen:2012ny}, 
the  low energy states of this system are fractional fillings of a $2N_5$-fold degenerate, charge neutral, fermionic 
Landau level.\footnote{This counting of the degeneracy
assumes that candidate ground states must be colour singlets, otherwise, there would be a further factor of $N$, 
the number of colour states of the fundamental representation fermion, in the degeneracy. 
With charge density $\rho$ and magnetic field $B$ (we always
assume $B>0$), we define the filling fraction as $\nu=\frac{2\pi \rho}{NB}$ as if the charge comes in
quanta of  $N$ and one Landau level is completely filled when $\rho=N\frac{B}{2\pi}$ and $\nu=1$. }  
The weak coupling argument for quantum Hall ferromagnetism can be applied and one would 
expect that the $2N_5$-fold degeneracy is  lifted and that incompressible, charge-gapped states appear
at filling fractions $\nu=0,\pm 1,\ldots,\pm N_5$.  Analysis
of the strong coupling limit using the string theory dual, the D3-probe-D5 brane system,  shows that, at least some of
these states with smaller values of $\nu$  are also there at strong coupling.     In the strong coupling states,   
the D5 branes blow up to form a D7 brane.  The D7 brane is capable of having incompressible integer Hall states
at non-zero values of the U(1) charge density.  
For large values of $N_5$, the phase diagram of this model was discussed in reference \cite{Kristjansen:2013hma}.

\begin{figure*}
\begin{center}
\includegraphics[scale=.45]{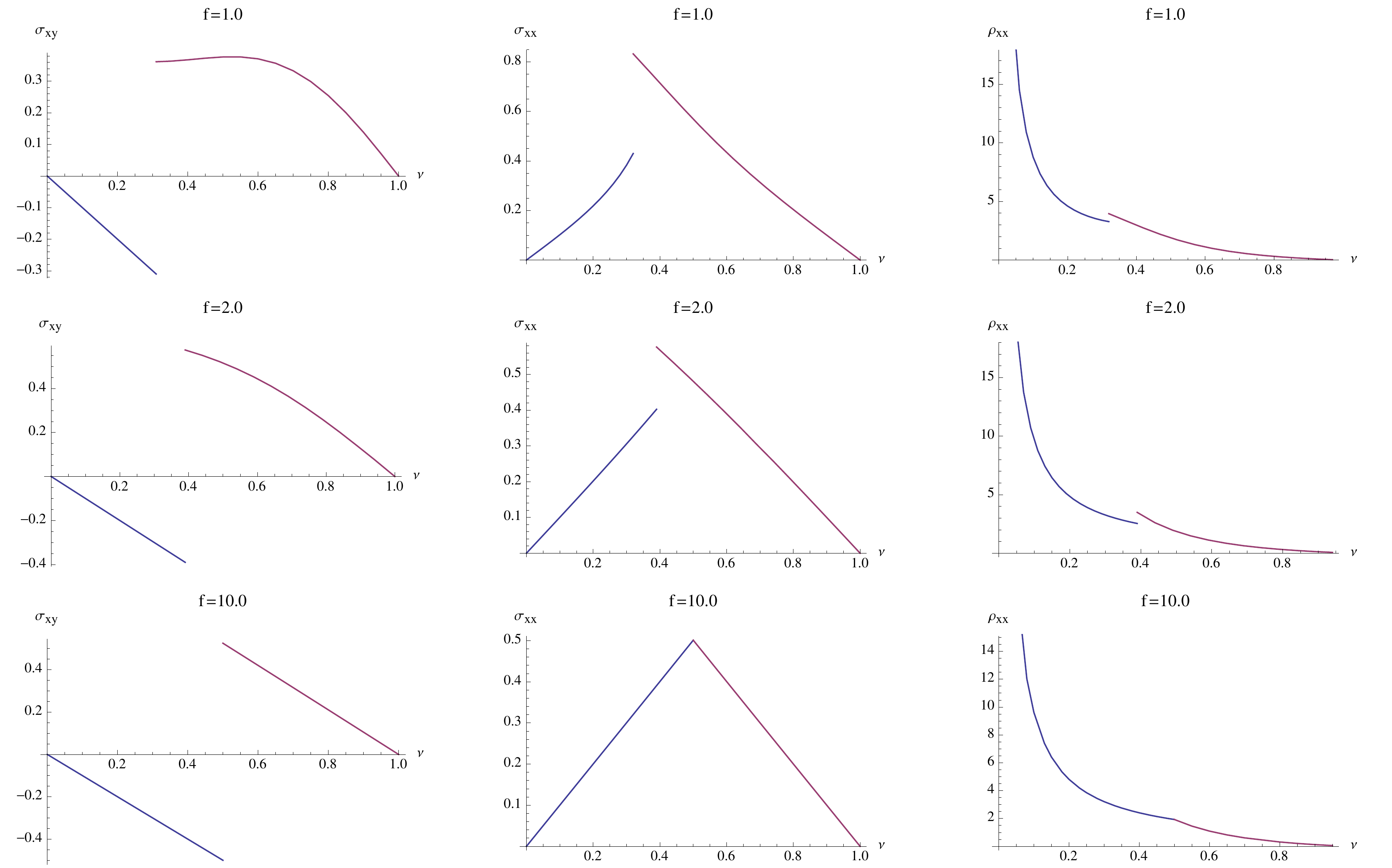}
\end{center}
\caption{\label{figure1} The first, second and third columns  are the deviation of the Hall conductivities $\sigma_{xy}$ 
from the classical Hall conductivity $\frac{N\nu}{2\pi}$,  the longitudinal conductivity $\sigma_{xx}$ and
 the longitudinal resistivity $\rho_{xx}$, respectively, for three different values of $f$ and for $\nu\in [0,1]$.
The temperature is such that $\hat r_h=0.2$  where $\hat r_h$ is defined in equation (\ref{definehatrh}).
The units of the $y$-axes are respectively
$ \hat r_h^4/(1+\hat r_h^4)\cdot\frac{N}{2\pi}$, $\hat r_h^2/(1+\hat r_h^4)\cdot\frac{N}{2\pi}$ and $\hat r_h^2\cdot\frac{2\pi}{N}$.}
\end{figure*}

\begin{figure*}
\begin{center}
\includegraphics[scale=.45]{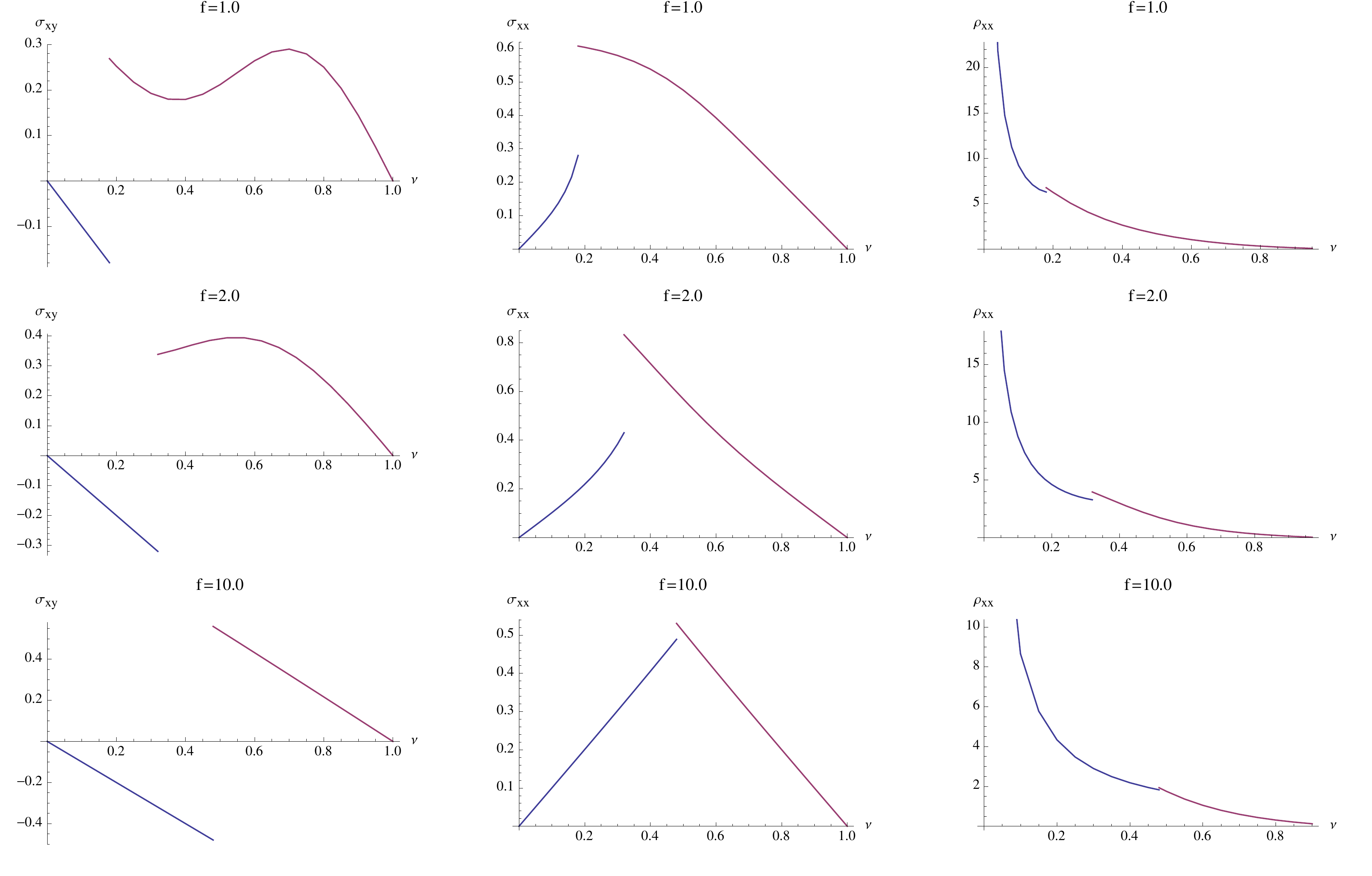}
\end{center}
\caption{\label{figure2}   The first, second and third columns  are the deviation of the Hall conductivities $\sigma_{xy}$ 
from the classical Hall conductivity $\frac{N\nu}{2\pi}$, the longitudinal conductivity $\sigma_{xx}$ and
the the longitudinal resistivity $\rho_{xx}$, respectively, for three different values of $f$ and for $\nu\in[0,1]$.
The temperature is such that $\hat r_h=0.4$.  The units of the $y$-axes are respectively
$ \hat r_h^4/(1+\hat r_h^4)\cdot\frac{N}{2\pi}$, $\hat r_h^2/(1+\hat r_h^4)\cdot\frac{N}{2\pi}$ and $\hat r_h^2\cdot\frac{2\pi}{N}$. }
\end{figure*}

In this paper, we shall compute the conductivity and the low temperature limit of the entropy 
of the strongly coupled states that are found in the D3-D5 model at finite temperature $T$,  density $\rho$ and
in a magnetic field $B$.  We concentrate on an interval of  
filling fractions between the integer quantum
Hall states, $0\leq \nu\leq 1$.  Our main aim is to explore the consequences of the phase transition from the
D5 to the D7 brane, 
which was found in references \cite{Kristjansen:2012ny}-\cite{Kristjansen:2013hma}, for the electronic transport
properties of the system.  
At the phase transition, which for the values of $f$ that we consider here, occurs at a critical value
of filling fraction $\nu_c\sim 0.3-0.5$, the stack of $N_5$ D5 branes, which are stable when $\nu<\nu_c$
blows up to a single D7 brane which is the preferred state when $\nu>\nu_c$.  
 The two   phases are distinguished
by their symmetry breaking patterns, U(N$_5$)$\times$SO(3)$\times$SO(3)$\to$U(N$_5$)$\times$SO(2)$\times$SO(3)
for the D5 branes  and U(N$_5$)$\times$SO(3)$\times$SO(3)$\to$U(1)$\times {\rm SO}(3)\times$SO(3)
for the D7 brane.    The D5 brane longitudinal conductivity, which we can find analytically in 
the  limit where 
the parameter $f=\frac{2\pi N_5}{\sqrt{\lambda}}$ is large, that is where  $N_5\gg\frac{\sqrt{\lambda}}{2\pi}$, 
is
\begin{equation}\sigma_{xx}^{D5}=\frac{     \frac{\pi\sqrt{\lambda} }{2B}T^2     }{1+\left( \frac{\pi\sqrt{\lambda} }{2B}T^2\right)^2 } \cdot \frac{N\nu}{2\pi} ~~,
\label{sigmaxxd5}
\end{equation}
This expression
 is a rather featureless 
linear function of the density, in particular, exhibiting no trace of the higher Landau level or
the insulating behaviour which should occur at integer quantum Hall states. This is remedied by
the phase transition. 
If we realize that, for large values of $f$, when $\nu=0.5$, the D5 branes are replaced by a D7 brane, 
the D7 brane conductivity should take over there.  In the large $f$ limit,  
\begin{equation}\label{sigmaxxd7} \sigma_{xx}^{D7}=  \frac{  \frac{\pi\sqrt{\lambda} }{2B}T^2    }{1+\left( \frac{\pi\sqrt{\lambda} }{2B}T^2\right)^2 } 
\cdot \frac{N~(1-\nu\,)}{2\pi} ~~,
\end{equation} 
a decreasing function of $\nu$ which reverts to an insulating state precisely when $\nu=1$. The result for
$\sigma_{xx}$
is depicted in the centre column
of figures \ref{figure1} and \ref{figure2} (for two different temperatures). 
What we find  for large values of $f$ (bottom row) is qualitatively like 
the longitudinal conductivity that would be expected to appear between integer Hall plateaux.
 The first and second
entries of the second columns
in  figures  \ref{figure1} and \ref{figure2}  show the behaviour for smaller values of $f$, 
where the conductivity is discontinuous at the phase transition.

\begin{figure}
\begin{center}
\includegraphics[scale=.4]{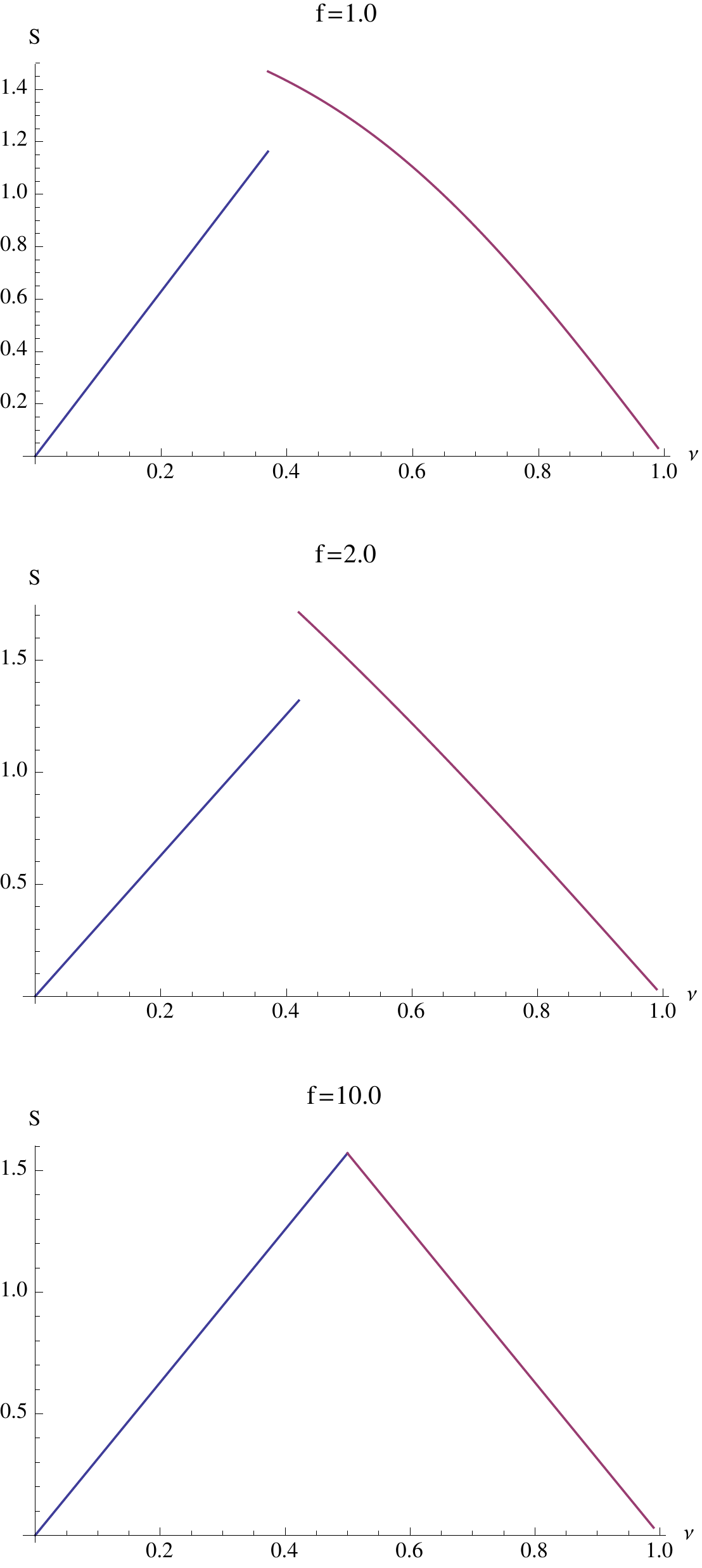}
\end{center}
\caption{\label{fig3}   The low temperature limit of the entropy density (in units of $\sqrt{\lambda}\cdot N\cdot B/(2\pi)^2$)
is plotted on the vertical axis versus filling fraction on the horizontal axis for various values of $f$. }
\end{figure}

The low temperature entropy exhibits similar behaviour.  If we first go to weak coupling and compute the zero temperature entropy of
the many-electron state coming from the degeneracy, $ \left(\begin{matrix} NBV/2\pi \cr NBV\nu/2\pi\end{matrix}\right)$,  of a partially filled Landau level,  
$$s_{\lambda\to 0}
\approx B\,\frac{N\nu}{2\pi}\ln\frac{1}{\nu}
+ B\,\frac{N(1-\nu)}{2\pi}\ln\frac{1}{1-\nu}.$$
Here, we have assumed that interactions have created the Hall ferromagnetic state, but are not strong enough to appreciably
 resolve the degeneracy of the partial fillings of the Landau level. To compare, we shall compute the low temperature entropy of the D5
brane (up to order $T^3$).  The result is identical to the one reported in \cite{Karch:2009eb}
\begin{equation}
s^{D5} = \frac{\sqrt{\lambda}}{2}\,B\, \frac{N\nu}{2\pi}~=~\frac{\sqrt{\lambda}}{2}\rho.~~
\label{sd5}
\end{equation}
Aside from the factor of $\frac{\sqrt{\lambda}}{2}$, which normally occurs in front of the entropy of a probe brane (see reference \cite{Karch:2009eb}
 for 
a discussion), this entropy increases linearly with the filling fraction.   Now, again, we realize that at a critical $\nu$,  
the D7 brane
takes over.  Our computation of the D7 brane entropy in the large $f$ regime gives
\begin{equation}
s^{D7} = \frac{\sqrt{\lambda}}{2}~B~ \frac{N(1-\nu)}{2\pi}.
\label{sd7}
\end{equation}
Interestingly, since $\nu_c=1/2$ in the large $f$ limit, this restores the $\nu\to 1-\nu$ symmetry of the weak coupling limit.  
The plot of low temperature entropy versus
$\nu$ for a few values of $f$ are displayed in figure \ref{fig3}. As in the case of the longitudinal conductivity, for finite $f$, they exhibit
discontinuities at the phase transition.

\begin{figure}
\begin{center}
\includegraphics[scale=0.55]{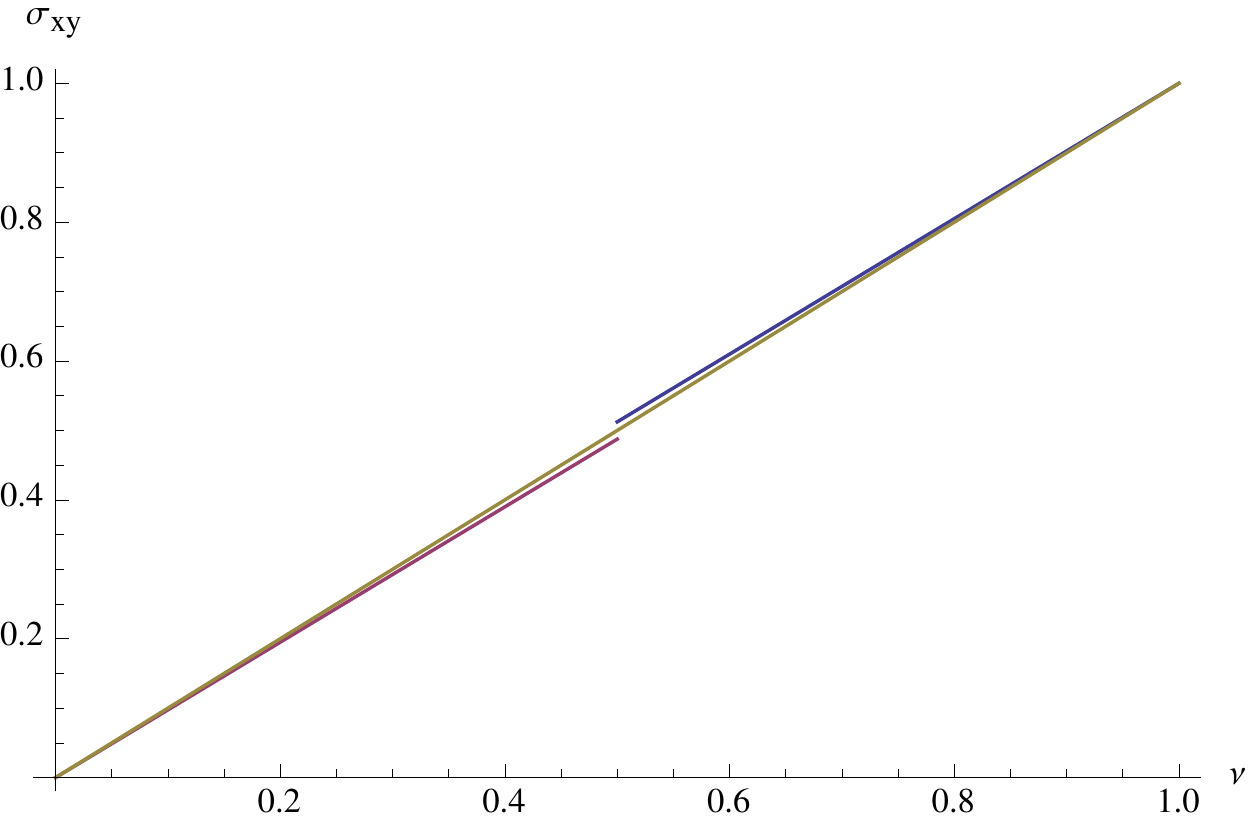}
\end{center}
\caption{\label{figure3}  The Hall conductivity for $f=10$ and $\hat r_h=0.4$ for 
$\nu\in [0,1]$ in units of $N/(2\pi)$. The red curve corresponds to the D5 brane
and the blue one to the D7 brane. For comparison we have also plotted the linear
curve (in green).
}
\end{figure}

The Hall conductivity does not exhibit integer Hall plateaux.  Of course, in the translationally  invariant system which we are considering here, the physics of
impurity driven localization which is normally responsible for Hall plateaux is absent. Moreover, in a Lorentz covariant system, there is 
an argument that the zero temperature Hall conductivity is identical to its classical value, $\sigma_{xy}=\frac{N\nu}{2\pi}$.  \footnote{The charge
density of a partially filled Landau level is $\rho = \frac{N\nu}{2\pi}B$.  We can create a constant current $J_i= \rho {\rm v}_i$ by going to a  reference frame
with velocity ${\rm v}_i$. 
The accompanying boost of the magnetic field creates a
 transverse electric field $E_i =-\epsilon_{ij}{\rm v}_j B$ 
 and we  have  $j_i = \frac{N\nu}{2\pi}\epsilon_{ij}E_j$ giving $\sigma_{xy}=\frac{N\nu}{2\pi}$. 
}  Our computation of the Hall conductivity at finite temperature nevertheless reveals an interesting dependence on $\nu$.
For example,  in the large $f$ limit, the Hall conductivities become
\begin{align}\label{sigmaxyd5}
\sigma_{xy}^{D5}&=\frac{N\nu}{2\pi} -   \frac{N\nu}{2\pi}\cdot \frac{\left( \frac{\pi\sqrt{\lambda} }{2B}T^2\right)^2 }{1+\left( \frac{\pi\sqrt{\lambda} }{2B}T^2\right)^2 } \,,  \\
\sigma_{xy}^{D7}  
&= \frac{N \nu}{2\pi}+ \frac{N (1-\nu)}{2\pi} \cdot \frac{\left( \frac{\pi\sqrt{\lambda} }{2B}T^2\right)^2 }{1+\left( \frac{\pi\sqrt{\lambda} }{2B}T^2\right)^2 }.
\label{sigmaxyd7}
\end{align}
At the zero temperature limit, the second terms in (\ref{sigmaxyd5}) and (\ref{sigmaxyd7}) vanish
and  the Hall conductivity is identical to the classical Hall value, $\lim_{T\to0}\sigma_{xy}=\frac{N\nu}{2\pi}$, as
expected. 
At finite temperature, 
the thermal correction decreases the  conductivity for the D5 and increases it for the D7 brane providing a jump
at the phase transition and a flattening of the slope of the $\sigma_{xy}$ versus $\nu$ curve.  If we could take the 
extreme high temperature
 limit, when $T^2\gg {2B/\pi\sqrt{\lambda}}$, in fact, $\sigma_{xy}^{D5}\to 0$ and $\sigma_{xy}^{D7}\to 1$ and we would 
 have perfect Hall plateaux with the Hall step occurring at the phase transition.  It is tantalizing to speculate that
 there is a strong coupling mechanism at play which, combined with temperature, gives a tendency toward plateau 
 formation.  However, in this system, we cannot take the large temperature limit.  We are limited to very low temperatures, 
 $\pi\sqrt{\lambda} T^2/2B< 0.16$, otherwise, the chiral symmetry is restored and there is no integer Hall state at all. As one can see
 in figure \ref{figure3}, the plateauing effect at this low temperature is miniscule. 
 Whether there exists an elaboration of our model where  the quantum Hall antiferromagnetic
 phase persists to higher temperature and the effect is more visible is an open  question which we shall not pursue in this paper. 
 The deviations of the Hall conductivities from the classical expression for three values of $f$ are displayed in the first 
 columns of figures \ref{figure1} and \ref{figure2}.

We will now outline our computation of the conductivity.  We follow a technique which was invented by Karch and O'Bannon~\cite{Karch:2007pd,O'Bannon:2008aj}.
Since there are several examples of how this technique is used in existing literature, we will be brief. 
We shall  work with the D3-D5 and D3-D7 probe brane systems studied in references~\cite{Kristjansen:2012ny,Kristjansen:2013hma} and follow   the notation of those references. One difference will be the use of  
Fefferman Graham rather
than Poincar\'{e} coordinates.  The   metric  of the AdS black hole is\footnote{The radial coordinate $z$
is related to our previously used one, $r$, by the equation $z^2=2 \left(r^2+\sqrt{r^4-r_h^4}\right)^{-1}$.  In particular, $z_h=\sqrt{2}/r_h$.}
\begin{align}\label{adsmetric}
ds^2_{AdS} = \sqrt{\lambda}\alpha'\left[ -\frac{(1-z^4/z_h^4)^2}{z^2(1+z^4/z_h^4)} dt^2+\frac{1+z^4/z_h^4}{z^2} 
d{\vec{x}}^{\,2} + \frac{dz^2}{z^2}\right],  
\end{align}
where $\vec{x}=(x,y,w)$.
The boundary of $AdS_5$ is located at $z=0$ and the horizon at $z_h$.  The Hawking temperature of the horizon is
$
 z_h =  \frac{\sqrt{2}}{\pi T}
$. 
We  parametrize $S^5$ as 
$$
ds^2_{S^5}= \sqrt{\lambda}\alpha'\left[d\psi^2+\sin^2\psi \,d\Omega_2^2+ \cos^2\psi \,d\tilde{\Omega}_2^2\right], $$
where 
$
d\Omega_2^2=d\theta^2+\sin^2\theta \,d\phi^2
$,  $ 
d\tilde\Omega_2^2=d\tilde\theta^2+\sin^2\tilde\theta \,d\tilde\phi^2
$.
 Our ansatz for the probe brane embedding is partially by symmetry.  
We will use world volume coordinates  $(t,x,y,z,\theta,\phi,\tilde{\theta},\tilde{\phi})$  for the 
the D7 brane with the ansatz that $w$ is constant
and $\psi=\psi(z)$ depends only on $z$. For the D5 brane we use coordinates $(t,x,y,z,\theta,\phi)$  where 
$\tilde{\theta}$,  $\tilde{\phi}$  are also constant.  The world volume metrics are 
\begin{align} ds_5^2
&= \sqrt{\lambda}\alpha'\left[ -g_{tt} \,dt^2 +g_{xx}\,(dx^2+dy^2)+g_{zz}\, dz^2+\sin^2\psi \,d \Omega_2^2
 \right],\label{D5metric}
 \nonumber \\ ds_7^2&=ds_5^2+\sqrt{\lambda}\alpha'\cos^2\psi \,d \tilde{\Omega}_2^2,
\nonumber \end{align}
where the metric components are 
\begin{align}
g_{tt}=\frac{1}{z^2}\frac{(1-\frac{z^4}{z_h^4})^2}{1+\frac{z^4}{z_h^4}} ,
\hspace{0.3cm} g_{xx}=\frac{1+\frac{z^4}{z_h^4}}{z^2},\hspace{0.3cm} g_{zz}=\frac{1}{z^2} +\left(\frac{d\psi}{dz}\right)^2.
\end{align}  
We make the ansatz for the world volume
gauge fields, 
\begin{eqnarray}
\lefteqn{2\pi\alpha'{\mathcal F}=\sqrt{\lambda}\alpha'\left[ \frac{d}{dz}a(z)dz\wedge dt + b\, dx\wedge dy  -e\, dt \wedge dx
 \right. \nonumber} \\ 
 &&\left. +\frac{d}{dz}f_x(z)dz\wedge dx  +\frac{d}{dz}f_y(z)dz\wedge dy+
\frac{f}{2}d\cos\tilde\theta\wedge d\tilde\phi
 \right].\nonumber 
\end{eqnarray}
with
\begin{equation}
B= \frac{\sqrt{\lambda}}{2\pi}~b, \hspace{0.5cm}
E=\frac{\sqrt{\lambda}}{2\pi} \, e,\hspace{0.5cm}
\rho =
\frac{1}{V_{2+1}}\frac{2\pi}{\sqrt{\lambda}}
\frac{\delta S}{\delta \frac{d}{dz}a(z)}\label{chargedensity},
\end{equation}
where $E$ and $B$ are constant external electric and magnetic fields, $\rho$ is charge density
and $V_{2+1}=\int dt \,dx\, dy $. The D5  and   D7 branes  have the same
values of $E$, $B$ and $\rho$.  We will  consider $N_5$ D5 branes but always a single D7 brane. The parameter 
$
f=\frac{2 \pi}{\sqrt{\lambda}} N_5
$ 
is proportional  to the
number of D5 branes and it becomes a world-volume flux on the the D7 brane.

The probe geometries are fixed once we find the functions $\psi(z),a(z),f_x(z),f_y(z)$.  These are determined
by requiring that they extremize the Dirac-Born Infeld (DBI) action with the addition of a Wess-Zumino (WZ) term
for the D5 or D7 brane. 
With the above ansatz, these actions take the form
\begin{align}
S_5&= -{\mathcal N}_5N_5\int_0^{z_h}  dz~  \sqrt{\cal{S}},          \label{S5}   \\
S_7&=-{\mathcal N}_7 \int_0^{z_h}  dz\left[~  (f^2+4\cos^4\psi)^{1/2}\sqrt{\cal{S}} 
\right.         \nonumber \\
&\left. 
+2 \left(a'(z)\, b\, -e\ f_y'(z)\right)c(\psi)\right] ,   \label{S7}
\end{align}
where $
{\mathcal N}_5=\frac{2\sqrt{\lambda}  N}{(2\pi)^3}V_{2+1},
\hspace{0.5cm}{\mathcal N}_7= \frac{2\lambda  N}{(2\pi)^4}V_{2+1},
$ and 
\begin{eqnarray}
\lefteqn{\hspace*{-0.3cm}{\cal S}=4\sin^4\psi[
g_{zz}\,(g_{tt}\,(b^2+g_{xx}^2)-g_{xx}\,e^2)
-a'(z)^2(b^2+g_{xx}^2)  \nonumber }\\
&&\hspace*{-0.3cm} + g_{tt}\,g_{xx}\, (f_x'(z)^2+f_y'(z)^2)
+2\, a'(z) \,e \, b \,f_y'(z) -e^2 f_y'(z)^2].\nonumber
\end{eqnarray}
The second term of eqn.~(\ref{S7}) is from the WZ term and 
\begin{align}
c(\psi)=\psi - \frac{1}{4}\sin4\psi -\frac{\pi}{2}. \label{cpsi}
\end{align}
The functions $a(z)$, $f_x(z)$ and $f_y(z)$ 
 are cyclic variables and they can be eliminated in terms of their conserved momenta $(q,q_x,q_y)$ which are
 defined   as 
$$
q\equiv \frac{1}{{\cal N}_{5,7}}\frac{\delta S_{5,7}}{\delta a'(z)},\hspace{0.5cm}
q_x\equiv -\frac{1}{{\cal N}_{5,7}}\frac{\delta S_{5,7}}{\delta f_x'(z)},\hspace{0.5cm}
q_y\equiv -\frac{1}{{\cal N}_{5,7}}\frac{\delta S_{5,7}}{\delta f_y'(z)}.
$$
The quantity $q$ is proportional to the charge density $\rho$ defined in (\ref{chargedensity}) and $(q_x,q_y)$ to the current densities $\left(J_x,J_y\right)$.
When the constants of integration $(q,q_x,q_y)$ are fixed, the appropriate energy functional is the Routhian
which is obtained from a Legendre transformation of the action. 
The Routhians are
$$
R_5=-N_5 {\cal N}_5 \int dz \, {\cal R} _5, \hspace{0.5cm}R_7=-{\cal N}_7\int dz \, {\cal R}_7,\nonumber 
$$
where
\begin{eqnarray}
\lefteqn{{\cal R}_{5}=\sqrt{g_{zz}} 
\left\{4\sin^4\psi
 \,(g_{tt}\,g_{xx}^2+b^2\,g_{tt}\,-e^2\,g_{xx})\nonumber\right. }\\
&&+g_{tt}\,q^2 -g_{xx}\,q_x^2
-g_{xx}\,q_y^2  \nonumber \\
&&\left.+\frac{2 b\, e \,q \,q_y}{g_{xx}}-\frac{b^2(q_x^2+q_y^2)}{g_{xx}}-e^2\left(\frac{q^2}{g_{xx}}-\frac{q_x^2}{g_{tt}}\right)  \nonumber 
\right\}^{1/2}.
\end{eqnarray}
The expression for ${\cal R}_7$ follows from ${\cal R}_5$ by replacing $4\sin^4\psi\to 4\sin^4\psi\,(f^2+4 \cos^4\psi)$, $q\to (q+2bc(\psi))$ and $q_y\to (q_y+2ec(\psi))$.  The  Karch-O'Bannon technique~\cite{Karch:2007pd,O'Bannon:2008aj} now finds a relationship between the current densities $(J_x,J_y)\sim (q_x,q_y)$ 
and the electric field by
requiring that the
world-volume is nonsingular.  
To this end, we rewrite ${\cal R}_{5,7}$ as\footnote{Note that the inverse of $\sqrt{g_{tt}}\sim (1-z^4/z_h^4)$ in ${\cal R}_{5,7}$   leads to a singularity in 
the integral over $z$ which must be done to find the free energy ${ R}_{5,7}$.  This singularity is not removed by the procedure which follows.  It is attributed to a limitation of the probe brane approximation
when computing the free energy.  See reference \cite{Karch:2008uy} for a discussion.}
$$
{\cal R}_{5,7}= \sqrt{\frac{g_{zz}}{g_{tt}}}\frac{1}{g_{xx}}\sqrt{B\cdot C - A^2},
$$
where in the case of ${\cal R}_5$
\begin{eqnarray}
A&=&q\, b\, g_{tt}-q_y\, e \, g_{xx}, \nonumber \\
B&= &g_{tt}\,g_{xx}^2+g_{tt} \,b^2-g_{xx}\, e^2, \nonumber \\
C&= &4\sin^4\psi\,g_{tt}\,g_{xx}^2+g_{tt}\,q^2-g_{xx}\,(q_x^2+q_y^2)\nonumber.
\end{eqnarray}
The expressions  for ${\cal R}_7$ follow by making the same replacements as explained just above.
 The expression $B$ is 
negative at the horizon, $z=z_h$, and positive at the asymptotic boundary of AdS, i.e for $z\rightarrow 0$. It must therefore have at least one zero
at some finite (positive) value of $z$, which we denote by $z^*$. Solving   $B=0$ one finds that there is only one
positive real root,  
\begin{eqnarray}
\lefteqn{\frac{z_*^4}{z_h^4}=\tilde{e}^2-\tilde{b}^2+\sqrt{(\tilde{e}^2-\tilde{b}^2)^2+2(\tilde{e}^2+\tilde{b}^2)+1}\nonumber} \\
&&-\sqrt{\left((\tilde{e}^2-\tilde{b}^2)+\sqrt{(\tilde{e}^2-\tilde{b}^2)^2+2(\tilde{e}^2+\tilde{b}^2)+1}
\right)^2-1}, \nonumber
\end{eqnarray}
where
$ 
\tilde{e}=\frac{z_h^2}{2}\,e$, $\tilde{b}=\frac{z_h^2}{2}\,b$.
We note that $z^*\rightarrow z_h$ as $e\rightarrow 0$. Like $B$,   
$C$ is negative at the horizon and positive at the boundary of AdS.
It must therefore also have a zero for a finite value of $z$ and, in order
for the Routhian to stay real, this zero must coincide with the one of $B$. Finally,
$A$ also has to vanish at the the common zero of $B$ and $C$. 
In summary $B(z=z^*)=0$ determines $z^*$.  Then $C(z=z^*)=0$ and $A(z=z^*)=0$ will determine $(q_x,q_y)$. This reasoning leads to
\begin{eqnarray}
q_y^{D5}&=&\frac{b \,q^{D5}}{b^2+g_{xx}^2(z^*)}\, e, \nonumber \\
q_x^{D5}&=& \frac{g_{xx}(z^*)\, e}{b^2+g_{xx}^2(z^*)} 
\sqrt{4 \sin^4\psi(z^*) (b^2+g_{xx}^2(z^*))+(q^{D5})^2}, \nonumber
\end{eqnarray}
and for the D7 brane
\begin{eqnarray}
q_y^{D7}&= &\left(\frac{ b\,q^{D7}\, -2\,c \,g_{xx}^2(z^*)}{b^2+g_{xx}^2(z^*)}\right) \,e,\nonumber \\
q_x^{D7} &=&\frac{g_{xx}(z^*)\, e}{b^2+g_{xx}^2(z^*)} \,  \times 
\left[\left(q^{D7}+2 b c(\psi(z^*))\right)^2+\right. \nonumber \\
&& \left. 
4 \sin^4\psi(z^*) (f^2+4\cos^4(\psi(z^*))(b^2+g_{xx}^2(z^*))\right]^{1/2},
\nonumber
\end{eqnarray}
where $g_{xx}^2(z^*)$ can be expressed as
\begin{equation}\label{gxx*}
g_{xx}^2(z^*)= \frac{2}{z_h^4}\left(1+\tilde{e}^2-\tilde{b}^2+
\sqrt{(\tilde{e}^2-\tilde{b}^2)^2+2(\tilde{e}^2+\tilde{b}^2)+1}\right). \nonumber
\end{equation}

We recall the normalizations  
$q^{D5}/b=\pi\nu/f$ and $q^{D7}/b=\pi\nu$.  
The conductivities are defined as 
$
\sigma_{xx}=\left. \frac{J_{x}}{E}\right|_{E=0},$ $\sigma_{xy}=\left. \frac{J_{y}}{E}\right|_{E=0}$ 
and, with the normalization of the currents, 
$
J_{x,y}^{D5}=\frac{ {\mathcal N}_5N_5}{V_{2+1}} \frac{2\pi}{\sqrt{\lambda}}\,q_{x,y}^{D5} \label{Jxy}$,
 $J_{x,y}^{D7}=\frac{ {\mathcal N}_7}{V_{2+1}} \frac{2\pi}{\sqrt{\lambda}}\,q_{x,y}^{D7} \label{Jxy}$
 we obtain the Hall conductivities
\begin{align}
\sigma_{xy}^{D5}&=  \left(1-\frac{{\hat r}_h^4}{1+{\hat r}_h^4}\right)\frac{N\nu}{2\pi} \,,    \nonumber 
\\
\sigma_{xy}^{D7}  
&= \frac{N \nu}{2\pi} \nonumber  \\
&+ \frac{N}{2\pi}\left( (1-\nu)+\frac{1}{2\pi}\sin 4\psi(z_h) -\frac{2}{\pi}\psi(z_h)\right) \frac{{\hat r}_h^4}{1+{\hat r}_h^4},
\nonumber
\end{align}
where the horizon radius is related to the Hawking temperature in units of inverse magnetic length,
\begin{align}
\hat r_h\equiv \frac{r_h}{\sqrt{b}}=\frac{\sqrt{2/b}}{z_h}=
  \sqrt{\frac{\pi\sqrt{\lambda} }{2B}}T,
\label{definehatrh}\end{align} 
and we are using natural units ($\hbar=c=k_B=1$). For these formulae to be valid, the temperature must be low
enough that the quantum Hall ferromagnetic phase is stable, that is, $\hat r_h<0.4$.
We have used  (\ref{cpsi}) as well as the fact that $z^*\rightarrow z_h$ as $e\rightarrow 0$. 
We see that the conductivities are completely determined by the value of the angle $\psi$ at the horizon when
the electric field is set to zero.  This angle depends on $f$ and the other parameters and must be
determined by numerical solution of the equation which determines $\psi(z)$.

For a given filling fraction $\nu$, 
the value of the Hall conductivity is larger for the D7 brane than for the D5 brane. 
(The two are equal only for the trivial solution $\psi=\pi/2$.) Hence, there is always an  upwards
jump in the Hall conductivity when the D5 brane ceases to be the favourable one and the D7 brane takes over, which for the
values of $f$ that we consider
occurs in the vicinity of $\nu\sim 0.3-0.5$.

 Finally, one finds numerically both for the D5 brane and the D7 brane that $f^2 \sin^4\psi(z_h)\rightarrow 0$ as
$f\rightarrow \infty$. 
 This allows us to find the limiting expression for the D7 brane Hall conductivity which we quoted in equation (\ref{sigmaxyd7}).
   The D5 brane Hall conductivity quoted in equation (\ref{sigmaxyd7})  is independent of $f$. For $f\rightarrow \infty$, the phase transition
   occurs at $\nu=1/2$. At the phase transition, where the D7 brane takes over from the D5 brane, there is a jump 
of $\frac{N}{2\pi} \,{\hat r}_h^4/(1+{\hat r}_h^4)$ in the
Hall conductivity.  

For the longitudinal conductivities we find
\begin{eqnarray}
\sigma_{xx}^{D5}&=&\frac{{\hat r}_h^2}{1+{\hat r}_h^4}  \frac{Nf}{2\pi^2}\nonumber 
\sqrt{4 \sin^4 \psi(z_h) (1+{\hat r}_h^4) +(\pi \nu/f)^2}, \label{longitudinalD5} \\
\sigma_{xx}^{D7}&=&\frac{{\hat r}_h^2}{1+{\hat r}_h^4} \, \frac{N}{2\pi^2} \times \nonumber  \\
&&\left[\left(\pi(1-\nu)-2\, \psi(z_h)+\frac{1}{2}\sin 4\psi(z_h)\right)^2
\right.
\nonumber \\
&&\left.+
4 \sin^4\psi(z_h) (f^2+4\cos^4(\psi(z_h))(1+{\hat r}_h^4)\right]^{1/2}.\nonumber
\end{eqnarray}
From here we find the analytic expressions  for the  limiting behaviours of the longitudinal conductivities quoted in 
equations (\ref{sigmaxxd5}) and (\ref{sigmaxxd7}).
Obviously, using our results we can also calculate the longitudinal resistivity 
$\rho_{xx}=\sigma_{xx}/(\sigma_{xx}^2+\sigma_{xy}^2)$. In the limit of large $f$ we find
\begin{eqnarray}
\rho_{xx}^{D5}&=&\hat r_h^2\,\left(\frac{N\nu}{2\pi}\right)^{-1}\hspace{0.5cm}\mbox{as}\hspace{0.5cm} f\rightarrow \infty, 
\nonumber\\
\rho_{xx}^{D7}&=&\hat r_h^2\,\left(\frac{N\nu}{2\pi}\right)^{-1} \left[\frac{\nu\,(1-\nu\,)}{\nu^2+\hat r_h^4}\right]
\hspace{0.5cm}\mbox{as}\hspace{0.5cm} f\rightarrow \infty. \nonumber
\end{eqnarray}

In order to find the  conductivity  when $f$ is finite, 
we need to know the embedding angle at the horizon, i.e.\ $\psi(z_h)$.
In reference~\cite{Kristjansen:2013hma} the equations of motion for   the D5 and the D7
probe branes were solved numerically for $E=0$ and $\psi(z_h)$ can be
extracted from that work. This allows us to compute $(\sigma_{xy},\sigma_{xx})$. 
In figure~\ref{figure1} and~\ref{figure2} we show the deviation of the Hall conductivities 
$\sigma_{xy}$  from the classical Hall conductivity $\frac{N\nu}{2\pi}$, 
the longitudinal conductivity $\sigma_{xx}$ and the the longitudinal resistivity $\rho_{xx}$, 
respectively, for three different values of $f$,   $f=1$, $f=2$ and $f=10$, and for $\nu\in[0,1]$. 
Figure~\ref{figure1} corresponds to $\hat r_h=0.2$ and figure~\ref{figure2} to $\hat r_h=0.4$.
For $\hat r_h=0.2$ and $f=10$ the curves are already indistinguishable from the corresponding
curves for $f\rightarrow \infty$. An interesting feature of the curves is that the
deviation of the Hall conductivity from its classical value qualitatively shows a behaviour
corresponding to the appearance of a Hall plateau. It is negative and decreasing 
for $\nu\in [0,\nu_c]$ and changes discontinuously at $\nu=\nu_c$ to becoming positive and
decreasing for $\nu\in [\nu_c,1]$ where 
$\nu_c\rightarrow 0.5$ as $f\rightarrow \infty$. In absolute value the 
 plateau is, however, not very pronounced. The size of the observed
quantum Hall effect is limited by the fact that
the temperature for which the probe-brane solutions are stable turns out to be dynamically
confined to $\hat r_h\leq 0.4$.
In figure~\ref{figure3} we show the
actual Hall conductivity of our model together with the classical linear curve.
All of the above pertains to the case $\nu\in [0,1]$. For $1<\nu<2$ the favoured system is a composite system consisting of a single gapped D7 brane and either a set of 
un-gapped D5 branes or a single un-gapped D7 brane with the former system being relevant for the smaller values of $\nu$ and the latter one for the larger values of $\nu$.  
The total flux $f_{tot}$ of the composite system must be distributed between the constituents in such a way that the energy is minimal. (A minimisation procedure to determine
the distribution of the flux was implemented numerically in reference~\cite{Kristjansen:2013hma}.)  
The conductivity of a composite system is the sum of the conductivities of its various constituents.  The Hall conductivity of the gapped D7 brane is independent of $f$ and so is that of the D5 branes.  The Hall conductivity of the single un-gapped D7 brane does depend (indirectly via $\psi(z_h)$) on the value of $f$ which due to the minimisation procedure
must be smaller than $f_{tot}$.  All this means that the deviation from the classical
Hall conductivity in the first part of the interval $\nu \in[1,2]$  looks as in the first
part of the interval $\nu\in [0,1]$  and in the second part of the interval $\nu\in [1,2]$ 
the deviation looks like
the deviation in the second part of the interval $\nu\in [0,1]$, but corresponding to a 
smaller value of the flux. This pattern continues as $\nu$ increases but the region where
D7+D5 is preferred over D7+D7 gets smaller and smaller.

The computation of the low temperature limit of the entropy is straightforward.  The procedure follows the technique
that is outlined in reference \cite{Karch:2009eb}. We consider the on-shell action, that is, ${ R}_5$ or ${  R}_7$ of the D5 or D7 brane, respectively.
The Routhian is the relevant thermodynamic potential when the total charge is fixed and we identify it with
the Helmholtz free energy.   Then, the entropy is defined as the negative of the partial derivative of the free energy by the 
temperature, and the entropy density is 
$$
s^{D5,D7}=- \frac{1}{V_{2+1}}\frac{\partial}{\partial T}{ R}_{5,7}.
$$
The procedure is easiest if one reverts to the Poincar\'{e} coordinates for $AdS_5$ which were used in references \cite{Kristjansen:2012ny},
\cite{Kristjansen:2013hma}.
Then, the essential observation is that, because their equations of motion depend on temperature only by terms with $T^4$, the 
derivative of the embedding functions must be at least of order $T^3$.  Similarly, the Routhian itself contains temperature only 
in terms of $T^4$ and its derivative is of order $T^3$.  Then, finally, the low temperature limit picks up the integrand evaluated on
the lower limit of the integral.  The result for the low temperature limit of the entropy is what is quoted in equations (\ref{sd5}) and
(\ref{sd7}) for large $f$ and displayed in figure \ref{fig3} for other values of $f$. 

We have computed the conductivity of the quantum Hall ferromagnetic states of the D3-D5 brane system and discussed some of 
the implications.   We have assumed that the stable phases are the homogeneous ones that were found in references \cite{Kristjansen:2012ny}
and
\cite{Kristjansen:2013hma}.   It is known that the probe brane systems, at least for some range of temperature and density, have instabilities
to forming inhomogeneous condensates \cite{Bergman:2011rf}-\cite{Jokela:2014dba}.
 Whether the system we have examined can have such instabilities and how they would affect the
electronic properties of the system is a fascinating subject which we
leave for further work.

\section*{Acknowledgments}
The work of G.W.S.~and J.H.~is supported in part by NSERC. G.W.S.~acknowledges the kind hospitality of 
the Niels Bohr Institute and NORDITA.
C.K.\ was supported by FNU through grant number DFF -- 1323 -- 00082. C.K.\ would like to thank the
University of British Columbia and NORDITA for their kind hospitality.





\end{document}